# Characterisation of the HollandPTC R&D proton beamline for physics and radiobiology studies

M. Rovituso[a,*], C.F. Groenendijk[b], E. van der Wal[a], W. van Burik[a], A. Ibrahimi[b,a], H. Rituerto Prieto[b,a], J.M.C. Brown[b,c], U. Weber[d], Y. Simeonov[e], M. Fontana[f], D. Lathouwers[b], M. van Vulpen[a] and M. Hoogeman[a]

[a]*Department of Research and Development, Holland Proton Therapy Centre Delft, Delft, The Netherlands*
[b]*Department of Radiation Science & Technology, Delft University of Technology, Delft, The Netherlands*
[c]*Optical Sciences Centre, Department of Physics and Astronomy, School of Science, Swinburne University of Technology, Hawthorn, Australia*
[d]*GSI, Helmholtzzentrum für Schwerionenforschung GmbH, Darmstadt, Germany*
[e]*University of Applied Sciences, Institute of Medical Physics and Radiation Protection, Giessen - Germany*
[f]*DE.TEC.TOR, Turin - Italy*

## Abstract
HollandPTC is an independent outpatient center for proton therapy, scientific research, and education. Patients with different types of cancer are treated with Intensity Modulated Proton Therapy (IMPT). In addition, the HollandPTC R&D consortium conducts scientific research into the added value and improvements of proton therapy. To this end, HollandPTC created clinical and pre-clinical research facilities including a versatile R&D proton beamline for various types of biologically and technologically oriented research. In this work, we present the characterization of the R&D proton beam line of HollandPTC. Its pencil beam mimics the one used for clinical IMPT, with energy ranging from 70 up to 240 MeV, which has been characterized in terms of shape, size, and intensity. For experiments that need a uniform field in depth and lateral directions, a dual ring passive scattering system has been designed, built, and characterized. With this system, field sizes between 2x2 cm$^2$ and 20x20 cm$^2$ with 98% uniformity are produced with dose rates ranging from 0.5 Gy/min up to 9 Gy/min. The unique and customized support of the dual ring system allows switching between a pencil beam and a large field in a very simple and fast way, making the HollandPTC R&D proton beam able to support a wide range of different applications.

*Corresponding author: m.rovituso@hollandptc.nl

## Introduction

Due to the physical properties of proton beams, proton radiotherapy has the potential to improve patient outcomes while minimizing side effects and increasing patient quality of life. However, the underlying biological mechanism is not fully understood, and proton therapy could also benefit from further technological advances. This has prompted the proton facilities to create a sustainable R&D environment where various types of *in vitro* and *in vivo* experiments can be conducted [1] along with advanced technological experiments aimed at developing new technology for proton therapy [2].

To this end, HollandPTC, which is an independent outpatient center for proton therapy, scientific research and education, has created clinical and pre-clinical research facilities including a versatile R&D proton beamline for various types of biological and technological research. HollandPTC is a ProBeam (Varian a Siemens Healthineers Company) isochronus cyclotron-based facility that features pencil beam in two clinical rotating gantries (equipped with scanning system), one fixed horizontal dedicated eye treatment room [3] and one fixed horizontal experimental room. The latter is connected to a biological laboratory for mainly cell culture work and work on tissue samples, to a physics and a chemistry laboratory for preparation of experimental setups and the work before and after irradiation, and to a radiobiological preparation room for both *in vitro* and *in vivo* experiments.

The broad range of experiments and associated requirements ask for beamline design that is versatile for pre-clinical experiments and can also be linked to clinical proton therapy. Our research beamline has a pencil beam which mimic the one used for clinical IMPT with an energy ranging from 70 up to 240 MeV. For experiments that need a uniform field in lateral direction, a dual ring passive scattering system that can produce uniform fields between 2x2 cm$^2$ and 20x20 cm$^2$ has been designed, built, and characterized. Moreover, to cover a target uniformly in depth, a 2D range-modulator [4] [5] is used, to create a Spread-Out Bragg peak of 25mm [6].

It is important that the beamline is well characterized such that the dosimetric conditions under which the experiments are conducted are known and the outcomes can be compared to experiments at other facilities. Therefore, the purpose of this paper is to describe the characteristics of the R&D proton beamline and familiarize potential users with its potential for biological and technological experiments.

## 2. Material and Methods

The experimental room of HollandPTC consists of a fixed horizontal proton beamline served by the ProBeam isochronus cyclotron which can deliver a therapeutic proton beam of energies between 70 and 240 MeV with beam currents at cyclotron extraction from 1 nA up to 800 nA. Figure 1 shows the layout and configuration of the experimental room. The end of the vacuum beam pipe consists of a Kapton exit window of 0.125 mm thickness after which a beamline target station with a length of 3.8 m is built in air. The latter is composed by modular tables of 750x750 mm$^2$ with an incorporated Thorlabs plate that can be removed and replaced within 1 mm precision with respect to the beam. A fixed room laser system defines one of the isocenter (reference point in the room where lasers cross) at 911 mm from the exit window. The laser system is extended along the whole target station to allow accurate alignment at different distances from the exit window. A beam monitor (BMI01, DE.TEC.TOR) with a 300 x 300 mm$^2$ sensitive area and Water Equivalent (WE) thickness of 0.6 mm is placed 100 mm behind the exit window. It consists of two planar integral chambers and is capable of online monitoring the proton beam in terms of beam intensity and time with a 1-ms resolution. Moreover, the BM01 is connected to the Varian ProBeam system as a trigger for a customized delivery system to control the irradiation in terms of delivery time or delivered protons.

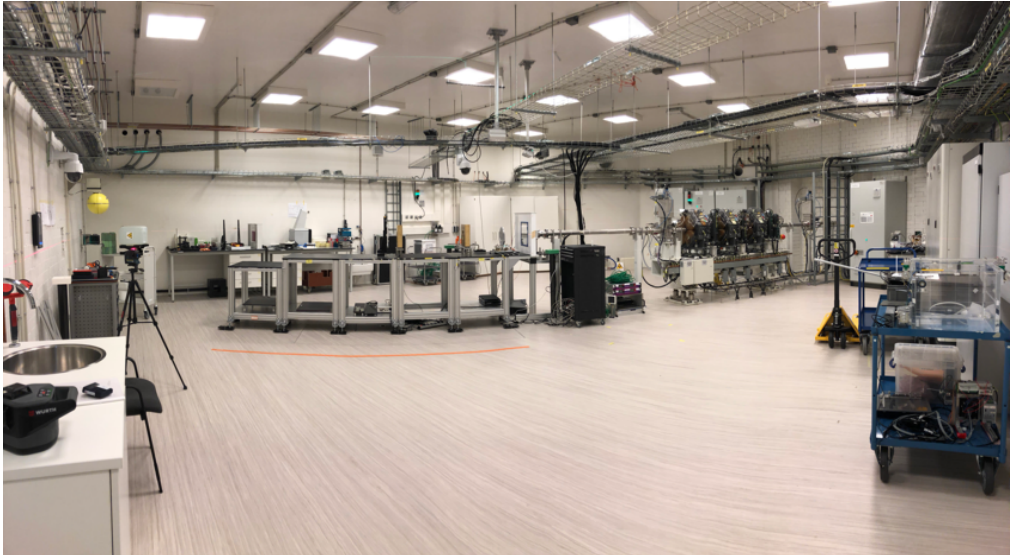
*Figure 1: Photo of the experimental room at HollandPTC*

*2.1 Pencil Beam*

To fully characterize the pencil beam, spot sizes, beam envelope, beam intensity, and depth dose distributions were measured. The following devices were used for the pencil beam characterization:
- The Lynx detector [7] (IBA Dosimetry, Schwarzenbruck, Germany): it consists of a scintillation screen coupled to a CCD camera and has an active surface area of 300x300 mm$^2$ with an effective spatial resolution of 0.5 mm in both *x* and *y* directions. The Lynx was used to measure the beam spot shape and size at isocenter and at different distances from the beam exit window for the beam envelope. It was also used to measure the large field shape and profile.
- BC-75 Faraday Cup (Pyramid Technical Consultants, Inc.): as a charge beam collector, it provides direct a measurement of the proton beam current optimized for proton energies up to 205 MeV, with a maximum current of 4uA. It has a sensitive area with a diameter of 70.5 mm and was used in this study as a reference to calibrate the beam monitor (BM01) and to characterize the beam current at isocenter.
- QubeNext (DE.TEC.TOR): it measures depth dose distributions through air vented multi-layer ionization chambers. It has a sensitive area of 127 x 127 mm$^2$ and contains 128 channels each corresponding to a pitch of 2.34 mm WE for a total proton range of 310 mm WE. The QubeNext is used to measure Depth Dose Distributions (DDDs).

The beam spot size for energies between 70 and 240 MeV was measured with the Lynx detector at isocenter. A 2D image was recorded and the projections in the *x* and *y* directions were extracted and analyzed. Both profiles are then modelled with a Gaussian function *f(x)*:

$$f(x) = A \cdot e^{-\frac{(\mu-x)^2}{\sigma}},$$

1)

where *A* is the height of the distribution, *μ* is the mean value and *σ* is the standard deviation of the measured distribution. The *μ* was used to determine the beam alignment with respect to the room laser system and the detector position, while the *σ* value was used to determine the beam spot size. For the latter the Full Width at Half Maximum (FWHM$_{x,y}$ = 2.355 · $\sigma_{x,y}$) was calculated both for *x* and *y*. The uncertainties of the determined beam spot size include the detector resolution (0.5 mm). The asymmetry (AS) of the beam spot was calculated as:

$$AS(\%) = 100 \cdot \frac{\sigma_x - \left(\frac{\sigma_x + \sigma_y}{2}\right)}{\frac{\sigma_x + \sigma_y}{2}},$$

2)

with $\sigma_x$ the beam spot size in the *x*-direction, and $\sigma_y$ the beam spot size in *y*-direction. The *x* and *y* beam profiles were measured in air at seven different positions from the exit window (230, 530, 911, 1230, 1530, 1830, 2045 mm) for 70, 120, 150, 200, and 240 MeV. The obtained beam envelope contains information on the beam focus, ion optics, scattering in air and beam alignment with respect to the fixed laser system. For each position in air and each energy, the *x-y* beam profiles were analyzed and modelled with the Gaussian function of Eq.1).

Because of the energy selection system of the cyclotron, the nominal beam intensity does not correspond to the beam intensity at room isocenter. Measurements to assess the transmission efficiency (TE) of the beam have been carried out using the Faraday cup BC-75. TE was measured for 10 different beam energies using a readout time of the BC-75 of 1 ms and readout range which varied from 1 nA to 10 nA. The TE is then defined as the ratio between the measured intensity at isocenter and the nominal beam intensity, as expressed in the following formula:

$$TE(\%) = 100 \cdot \left(\frac{I_{iso}}{I_{nominal}}\right).$$

3)

For the BM01 to be fully integrated in the beam delivery system, a calibration of the raw counts with respect to the number of protons was carried out using the BC-75. The latter was placed at isocenter behind the BM01 so that both devices were recorded simultaneously with a 100 msec readout time. For each energy, the irradiation was repeated 3 times with a duration of 20 seconds and a nominal beam intensity of 50nA. The calibration factors (CF) of BM01 for each energy, were defined as:

$$CF = \frac{N_{protons}}{N_{counts}}.$$

4)

Depth dose distributions of energies between 70 and 240 MeV were measured using the QUBEnext detector with steps of 10 MeV. The distributions were fitted with a Bortfeld function [8] and peak-to-plateau ratio, peak width, and the R80 (the position in depth where the beam is at 80% of the dose with respect to the maximum value) were extracted.

*2.2 Large passively scattered field*

Two dual ring passive scattering systems [9] [10] were built and optimized for an initial energy of 150 MeV to produce large fields of different sizes with a uniform dose distribution. One dual ring system was produced for fields up to 20 cm diameter with low dose rates, referred to as thick ring, and the other one for fields up to 8 cm diameter with higher dose rates, referred to as thin ring. The dual-ring system is composed by a simple lead foil and a ring with aluminum and lead. The thicknesses of the lead foil is 0.8 mm and 3.4 mm for thin and thick ring, respectively. The thickness and dimension of the rings are reported in Table 1. These dimensions were taken from the work of Tommasino et al. [11]. The dual-ring setups have been mounted on a special support, which allows them to be inserted and removed from the beam line without compromising the alignment with respect to the beam. The support of each ring is made in such a way that precise adjustment for alignment (in the order of 0.05 mm steps) can be done if necessary. A photo of the mounting is shown in Figure 2.

*Table 1: Dimensions of the dual-ring setup*

| Setup | Aluminum (Al) | | Lead (Pb) | |
|---|---|---|---|---|
| | Diameter (mm) | Thickness (mm) | Diameter (mm) | Thickness (mm) |
| Thick dual ring | 200 | 16 | 45 | 5.5 |
| Thin dual ring | 200 | 5 | 11 | 1.5 |

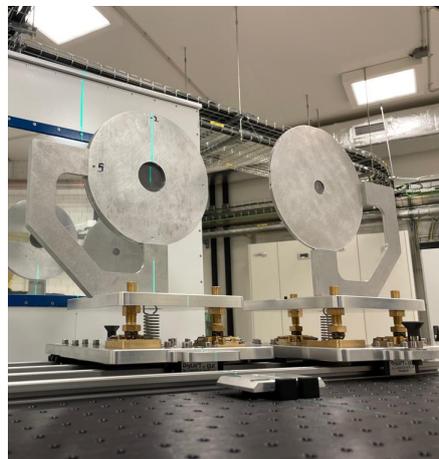

*Figure 2: Photo of the dual-ring setups mounted on a customized support*

In order to spread the monoenergetic beam in the longitudinal direction, a 2D range modulator [4] [5] is used. The 2D range modulator is also optimized for an initial energy of 150 MeV to produce a SOBP of 25 mm.

Figure 3 shows a photo and the schematic representation of the beam line elements of the dual ring passive scattering system. The proton beam exiting the vacuum exit window (1), passes through the first scattering foil of lead (2) which produces the initial beam lateral spread. The BM01 is placed upstream to monitor the number of protons delivered. The beam will then traverse the dual ring scattering (4) composed of an outer ring of aluminum and an inner ring of lead. The latter will offer an increase of the lateral spread of the beam produced by the first foil with a homogeneous intensity. The inner ring produces a Gaussian-like profile, while the outer ring produces an annulus-shaped profile. The passively scattered beam will then interact with the 2D range modulator (5) which will spread the beam longitudinally. The Brass collimator system (6) will restrain the field to different sizes and shape the distal edge of the field. A 2-stage Brass collimator is used in order to reduce activation close to the target [12]. Squared shape field sizes of 2,4,10, 16, and 20 cm are produced.

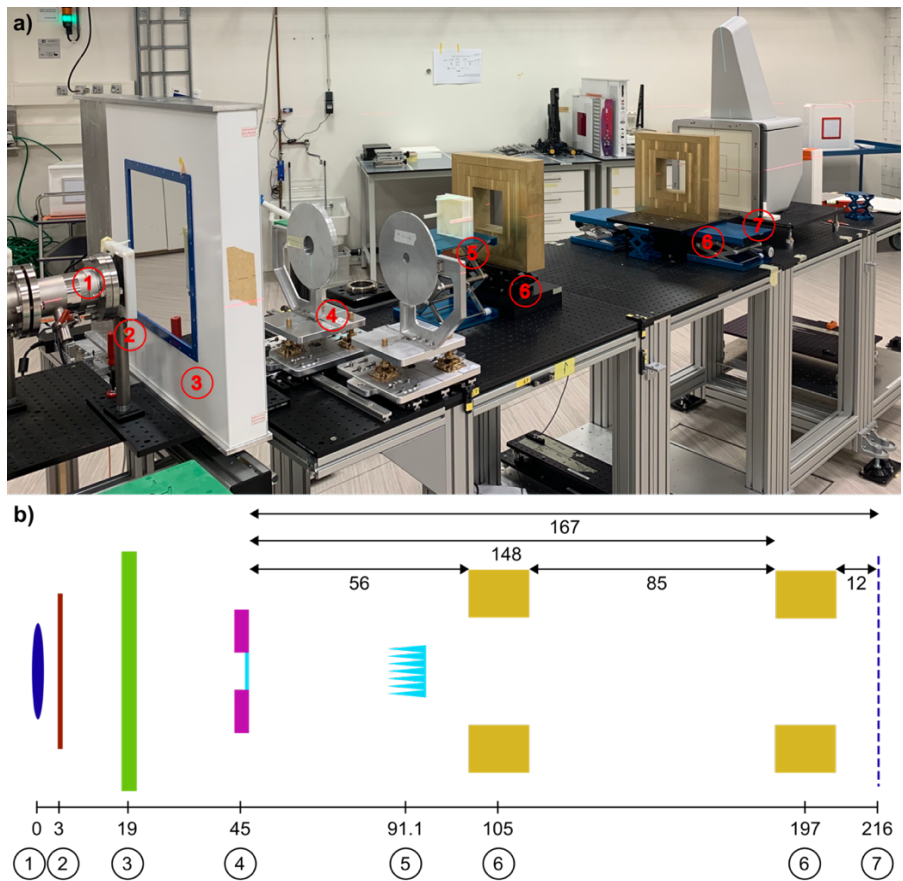

*Figure 3: (a) Photo of the experimental setup. (b) Schematic representation (not to scale) of the beam line elements with the distances from beam vacuum exit window. From left to right: the beam vacuum exit window, the first scattering foil, BM01, dual ring, 2D range modulator, collimator system and detector or target.*

To achieve maximum beam uniformity in the field, alignment between the beam and the dual ring scatterer is critical. Moreover, the distance between the first foil and the dual ring with respect to the target position must be optimized. For this purpose,

Geant4-TOPAS simulations have been carried out to find the optimal distance between first foil and dual ring and target position. For the alignment of the dual ring with respect to the beam in the *x-y* direction a customized system with motorized screws was built to support the dual ring and obtain an alignment within 0.1 mm. Once the optimal settings were found, the dual ring system was attached to a fixed support. The latter was built onto on a rail system, which allows the system to slide in and out the beam path, without having any impact on the alignment with respect to the beam. The shape and uniformity of the passively scattered field were measured using the Lynx detector. The Lynx was placed at target position, as shown in Figure 3. A 2D image of the beam is recorded and the profiles in *x* and *y* directions were extracted. The uniformity (*U*) of each field can then be calculated as

$$U(\%) = \left(1 - \frac{I_{max} - I_{min}}{I_{max} + I_{min}}\right) \cdot 100\%,$$

5)

where $I_{max}$ is the maximum intensity across the field of interest, and $I_{min}$ is the minimum intensity. A uniformity above 97% needs to be achieved for radiobiological experiments.

Once the uniformity of the field was assessed, the DDD of both Thin and Thick ring was measured in the water phantom with a PTW Advanced Markus® Electron Chamber. Moreover, the latter was used to perform absolute dose measurements. The Advanced Markus chamber used in the R&D room was cross-checked with the one used for clinical practice, which is calibrated against a primary standard. An agreement of 0.5% was found with the clinically used Advanced Markus chamber. To check absolute dose uniformity over the field length, measurements with the Advanced Markus chamber were performed in the middle of the field and along a few points in *x* and *y* direction. The dose rate measurements were always performed at the depth of the experiment (entrance BP channel, middle of SOBP, etc.).

To fully characterize the total dose distribution laterally and longitudinally, the OCTAVIUS Detector 1500XDR, was used. This detector consists of a matrix of vented plane-parallel ion chambers of 4.4 x 4.4 x 3 mm$^3$ in size and a center-to-center spacing of 7.1 mm. In total there are 1405 ion chambers arranged in a chessboard matrix, providing a maximum field size of 27 x 27 cm$^2$. Slabs of RW3 have been placed in front of the detector and *x-y* profiles in dose were measured at every depth. The plateau region was resolved with bigger steps in depth (1 cm), while in the SOBP region 5-mm steps were used. The 3D dose distribution was then reconstructed by integrating the *x* and *y* profiles at each depth.

*2.3 Target station for in-vitro radiobiological experiments*

In the R&D proton beam line of HollandPTC, clinically relevant conditions for in-vitro experiments have been realized. A dedicated flipper system for cell irradiation has been designed and built in-house, which allows cells to lay horizontally till few seconds before being irradiated. The technical drawings are presented in Figure 4

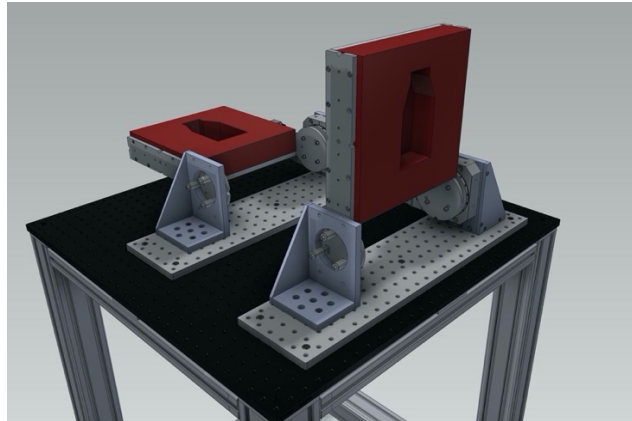

*Figure 4: Technical 3D drawings of the flipper system realized for the HPTC proton beam line. The sample holder shown in the drawings is for a T25 flask. In the figure both the vertical and horizontal positions are shown.*

A remote-controlled motor will flip the culture flask up to a vertical position when the beam is ready to be delivered. After the irradiation is completed, the flask will be flipped down into the horizontal position again. With this system the cells can be irradiated being immersed or not in culture medium with no impact from a biological point of view. 3D printed support for T25 culture flasks, well-plate, and petri-dish have been realized to irradiate the samples with this flipper system.

The desired depth along the BP or SOBP is realized using RW3 phantom slabs, which have thickness of 1 mm, 2 mm, 5 mm, and 10 mm. However, for certain type of experiments, it is best to replace the RW3 with water. For this purpose, a water phantom system was built to submerge the cell samples (T25 flask, well-plate) into water for irradiation. 3D printed supports were produced to mount the sample on a motorized linear stage which can move into the desired depth along the BP (or the SOBP). Moreover, the holder contains a slot for the Advanced Markus chamber positioned right behind the cell sample which can be recorded also during the irradiation.

BP and SOBP curve have been measured with the Advanced Markus chamber with and without the cell sample in front of it to assess the corrected water equivalent thickness of the cell sample. In Figure 5 the setup for a T25 flask holder is shown.

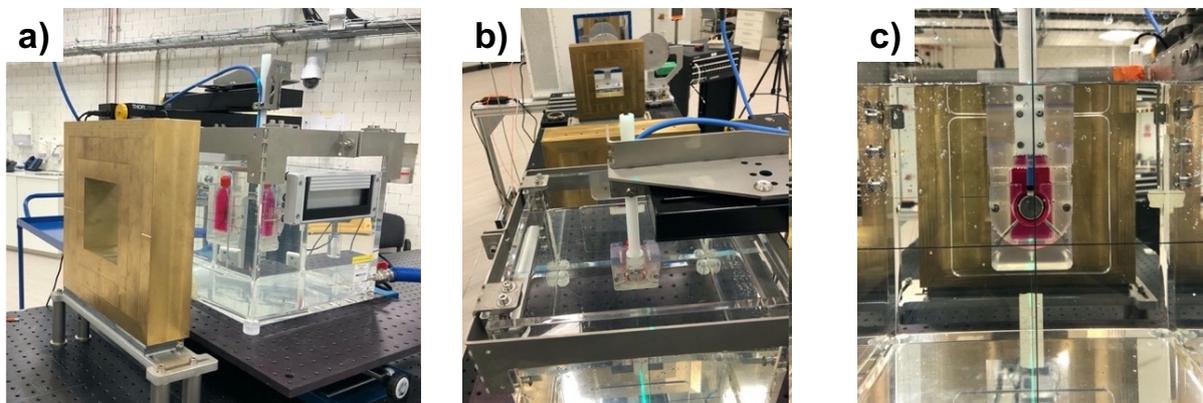

*Figure 5: Photos of the water phantom setup used for absolute dosimetry in water and for cell irradiation. A flask with culture medium is placed in front of the Advanced Markus chamber. a): a side view of the water phantom placed in front of a 10x10cm$^2$ collimator; b) a top view and c) a back view where the Advanced Markus chamber is visible.*

## 3. Results

*3.1 Pencil beam*

The beam spot sizes measured with the Lynx detector for energies between 70 and 240 MeV with 10 MeV steps are shown in Table 2 and Figure 6.

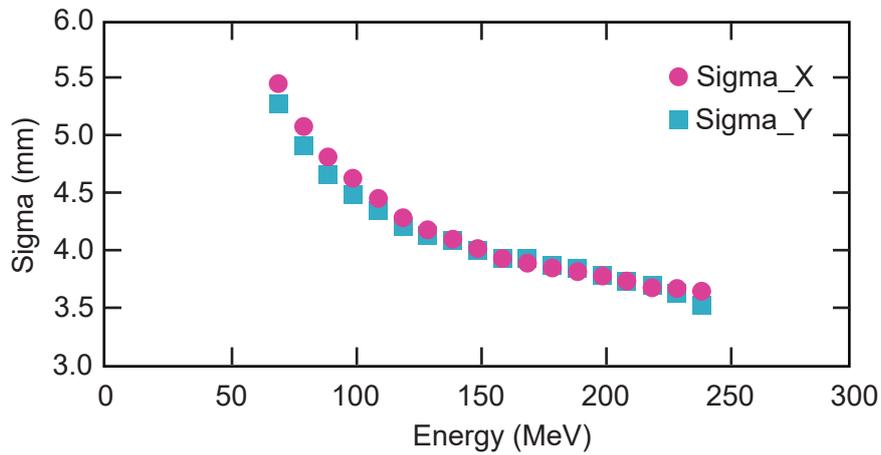

*Figure 6: Plot of the σ values in mm, extracted by the Gaussian fit function of the x and y beam profiles as function of nominal beam energy E(MeV).*

The experimental data could be reproduced with a Gaussian function Eq.1) and *σ* values have been extracted both for *x* and *y* beam profiles. Figure 7 shows the profile in *x* direction for 70, 150 and 240 MeV.

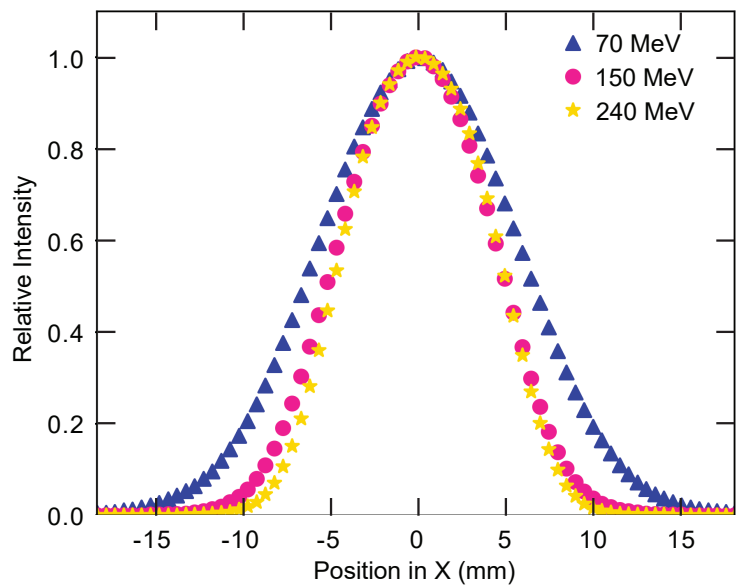

*Figure 7: Lateral profiles at isocenter for the x beam direction at 70, 150 and 240 MeV*

Beam size decreases with increasing beam energy, due to the cyclotron energy selection system (ESS) and the less pronounced multiple Coulomb scattering which higher energy protons undergo in air. Results of the beam envelope measurements for 70, 120, 150, 200 and 240 MeV are shown in Figure 8.

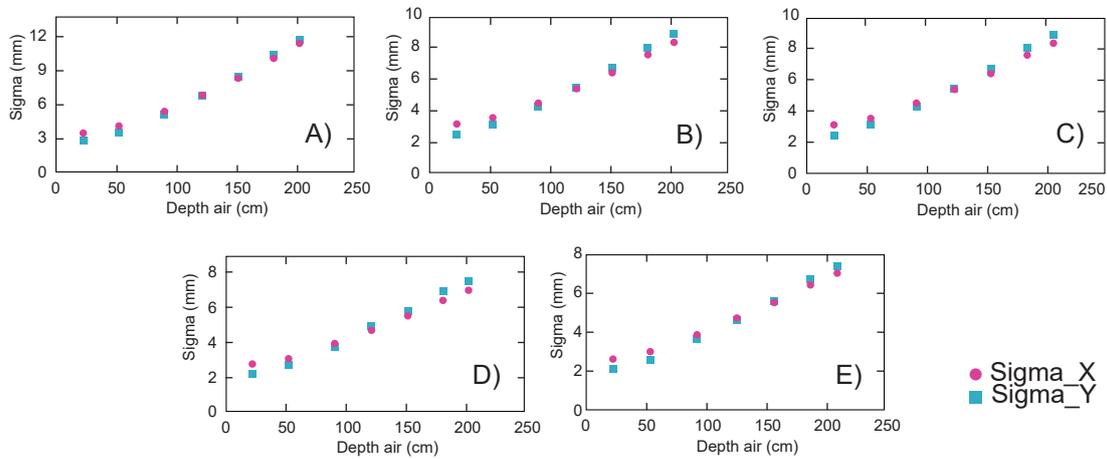

*Figure 8: Beam envelope experimental data. The σ values extracted from the Gaussian fit expressed in mm are shown as function of depth in air (cm) for 70, 120, 150, 200, and 240 MeV. The position 0 in depth refers to the vacuum exit window.*

A convergent beam around isocenter position (91cm from exit window) is shown for all energies. The AS was calculated at each position from exit window to show the effect of the ion beam optics on the beam shape and results are shown in Figure 9.

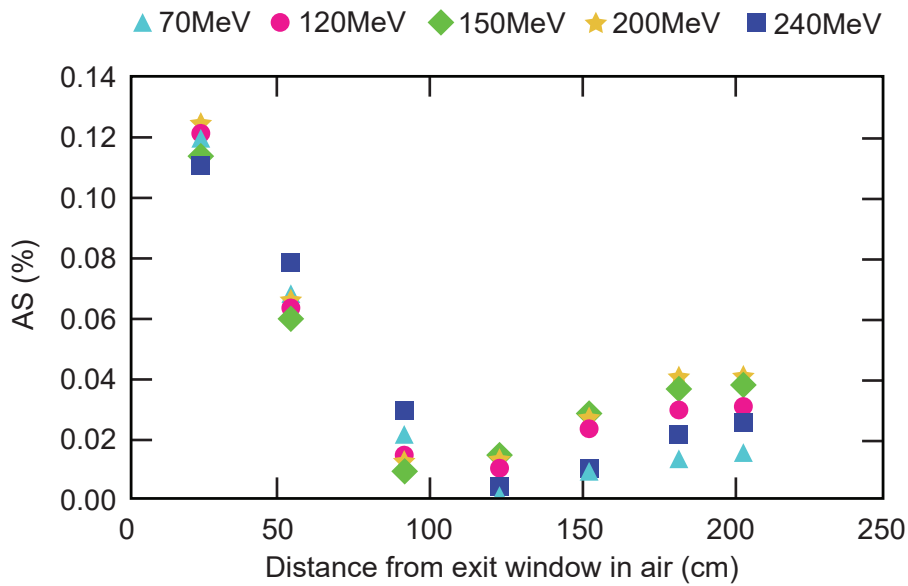

*Figure 9: Beam spot asymmetry (AS) calculated for each position along the beam direction from vacuum exit window up to 230 cm in air.*

The strongest AS is shown in front of the isocenter (focus point of the ion optics), increasing up to 4% at 205 cm. AS values at isocenter position are reported in Table 2. Furthermore, Table 2 shows the transmission efficiency (TE) of each energy and the number of protons per second per nominal current in nA, measured with the Faraday Cup BC-75. As expected, TE increases with increasing energy because of the cyclotron ESS.

The DDDs obtained with the QUBEnext detector for energies between 70 and 240 MeV with step of 10 MeV are shown in Figure 10. The effective energy at isocenter position with respect to the nominal energy at cyclotron extraction is reported in the second column of Table 2. Moreover, for 70, 100, 130, 150, 180, 200, and 240 MeV, the peak position, the peak width, the R80 and the peak to plateau ratio have been extracted modelling the experimental data with a Bortfeld function [8]. The results are reported in Table 3. All values in the DDDs plots and tables are expressed in cm Water Equivalent (cm WE).

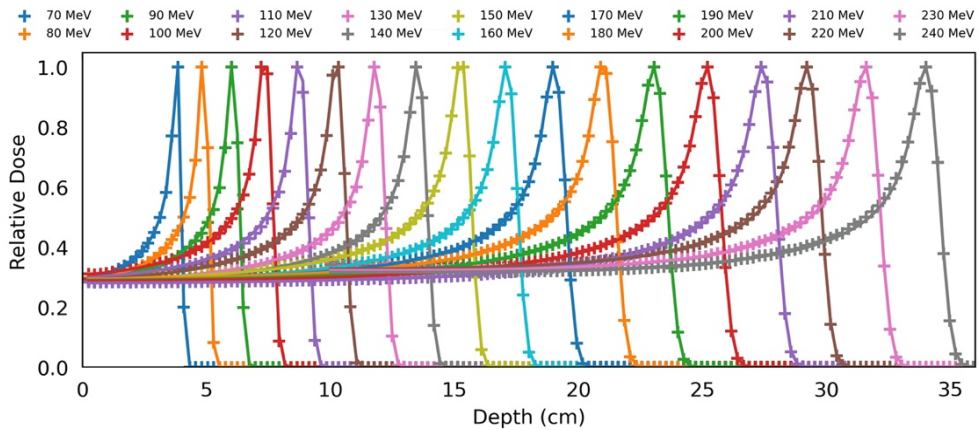

*Figure 10: Depth dose distributions obtained with the QubeNext detector for energies between 70 and 240 MeV with step of 10 MeV, measured at isocenter.*

*Table 2: Values of nominal beam energy $E_{nominal}$ in MeV, measured energy at isocenter $E_{iso}$ in MeV, beam spot size estimated from the Gaussian fit on the x-y profiles, beam spot asymmetry (AS) calculated at isocenter, trasmission efficiency (TE) and measured number of protons per second per nominal 1nA.*

| $E_{nominal}$ | $E_{iso}$ | $\sigma_x$ (mm) | $\sigma_y$ (mm) | AS (%) | TE (%) | Flux (p/s) |
|---|---|---|---|---|---|---|
| 70 | 69.2 | 5.47 | 5.29 | 1.6 | 0.04 | $2.44 \times 10^6$ |
| 80 | 79.4 | 5.10 | 4.93 | 1.6 | | |
| 90 | 89.6 | 4.83 | 4.67 | 1.7 | 0.07 | $4.44 \times 10^6$ |
| 100 | 99.8 | 4.64 | 4.50 | 1.5 | 0.11 | $6.86 \times 10^6$ |
| 110 | 109.9 | 4.47 | 4.36 | 1.2 | | |
| 120 | 120.0 | 4.30 | 4.23 | 0.8 | 0.14 | $8.91 \times 10^6$ |
| 130 | 129.9 | 4.19 | 4.15 | 0.5 | | |
| 140 | 139.9 | 4.10 | 4.12 | 0.2 | | |
| 150 | 149.7 | 4.02 | 4.04 | 0.3 | 0.29 | $1.78 \times 10^7$ |
| 160 | 159.6 | 3.95 | 3.95 | 0 | | |
| 170 | 169.3 | 3.91 | 3.94 | 0.5 | | |
| 180 | 179.0 | 3.86 | 3.90 | 0.4 | 0.63 | $3.94 \times 10^7$ |
| 190 | 188.6 | 3.83 | 3.86 | 0.4 | | |
| 200 | 198.2 | 3.80 | 3.81 | 0.1 | 1.13 | $7.04 \times 10^7$ |
| 210 | 207.8 | 3.75 | 3.75 | 0 | | |
| 220 | 215.3 | 3.72 | 3.70 | 0.3 | 1.87 | $1.16 \times 10^8$ |
| 230 | 224.8 | 3.69 | 3.63 | 0.7 | | |
| 240 | 237.1 | 3.66 | 3.54 | 1.6 | 5.27 | $3.29 \times 10^8$ |

*3.2 Large passively scattered field*

The passive scattering system of the R&D beam line of HollandPTC provides field sizes from 2x2 cm$^2$ up to 20x20 cm$^2$ with high and low dose rate. Figure 11 shows the 2D profiles of low dose rate field size of 4x4 cm$^2$, 10x10 cm$^2$ and 20x20 cm$^2$. All the produced fields with thick and thin ring have uniformity between 97% and 99%. Table 4 shows the uniformity of all available fields.

*Table 3: Parameters derived from the experimental data modelled with a Bortfeld function: peak position expressed in cm WE, peak width at R80 in cm WE and peak-to-entrance plateau ratio, for 70, 100, 130, 150, 180, 200, 240 MeV*

| E [MeV] | peak position (WE cm) | peak width (WE cm) | R80 [WE cm] | P-E ratio |
|---|---|---|---|---|
| 70 | 3.62 | 0.41 | 3.76 | 2.34 |
| 100 | 7.29 | 0.55 | 7.42 | 2.59 |
| 130 | 11.66 | 0.75 | 12.05 | 2.64 |
| 150 | 15.04 | 0.91 | 15.52 | 2.56 |
| 180 | 20.87 | 1.17 | 21.39 | 2.42 |
| 200 | 25.25 | 1.30 | 25.68 | 2.42 |
| 240 | 34.28 | 1.32 | 34.79 | 2.41 |

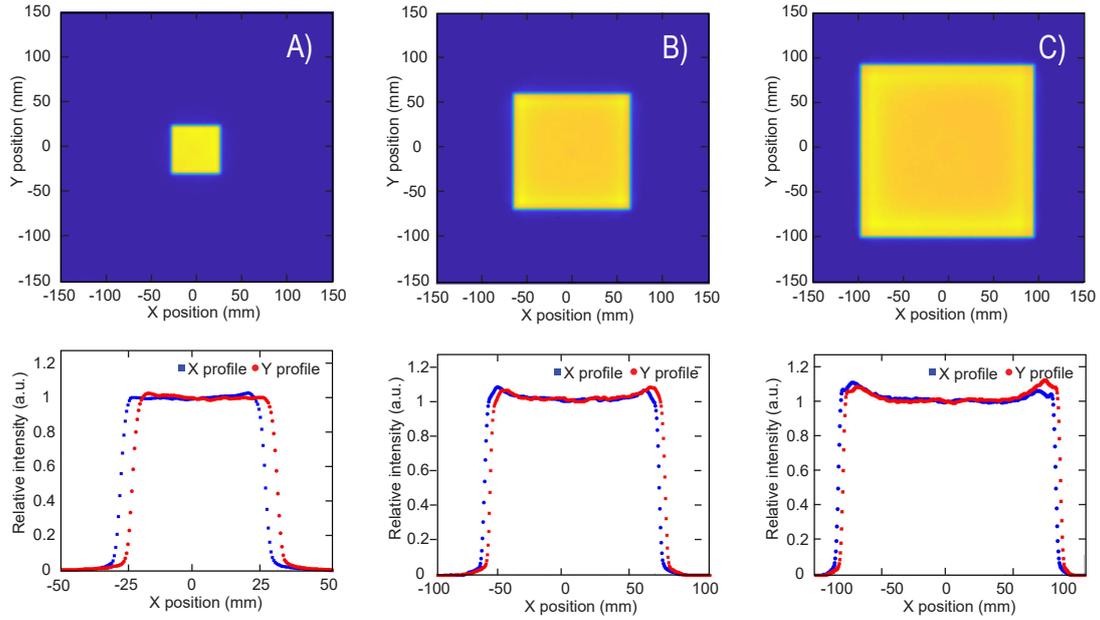

*Figure 11: 2D images of the large passively scatter field and the extracted lateral profiles in x and y. Field sizes of 4x4, 10x10 and 20x20 cm² obtained with the thick ring setup are shown. The intensity of the profiles is normalized to the maximum value.*

*Table 4: Summary of the available collimated field sizes and their uniformity U*

| Setup | Field size (cm$^2$) | Uniformity (%) |
|---|---|---|
| Thick Ring (low dose rate) | 2 × 2 | 99±1 |
| | 4 × 4 | 99±1 |
| | 10 × 10 | 98±1 |
| | 16 × 16 | 97±1 |
| | 20 × 20 | 97±1 |
| Thin Ring (high dose rate) | 2 × 2 | 99±1 |
| | 4 × 4 | 98±1 |
| | 8 × 8 | 98±1 |

The depth dose distribution of 150 MeV proton beam traversing the dual ring passive scattering system and the 2D range modulator, has been measured with the OCTAVIUS Detector 1500XDR. RW3 plastic slabs have been inserted in front of the detector and the *x* and *y* profiles measured in dose. The Bragg curve and the SOBP curve are shown in Figure 12 both for the thin and the thick ring setup.

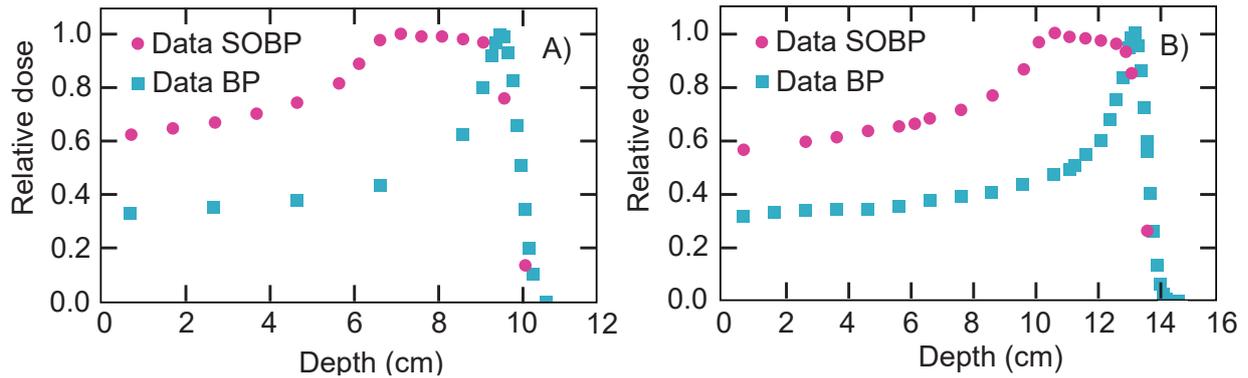

*Figure 12: Bragg peak and Spread-out-Bragg peak measured with Octavius 1500XDR in RW3 phantom for the thick ring (a) and the thin ring setup (b) are shown.*

The middle of the SOBP is found at 12.2 cm and 7.5 cm for the thin ring and thick ring, respectively. The ratio between SOBP value and entrance value is 1.57 for the thick ring and 1.74 for the thin ring. The width of the SOBP is 20 mm for the thin ring with a uniformity of 98% and 25 mm for the thick ring with a uniformity of 99%. The pristine Bragg peak of the two passive scattering setups without the 2D range modulator was measured in the same way. The R80 was found to be at 13.79 cm WE and 9.95 cm WE for thin and thick ring, respectively. The peak to entrance ratio was found to be approximately 3 for both setups. The maximum dose rates achievable for both setups were measured with the Advanced Markus chamber at the maximum cyclotron current at the middle of the SOBP. A dose rate of 9.9 Gy/min for the thin ring, while a dose rate of 1.6 Gy/min has been found for the thick ring. Maximum dose rate measurements have been performed also along the pristine Bragg peak to be able to perform experiments at different depths. For the thin ring dose rates of 5 Gy/min, 13 Gy/min, and 4 Gy/min have been found at the entrance, proximal 80 and R20 positions respectively. For the thick ring, instead, dose rates of 0.9 Gy/min, 2 Gy/min and 0.5 Gy/min have been found at the entrance, proximal 80 and R20 positions respectively. 2D dose reconstructions have been performed with the experimental data of the Octavius 1500 XDR in the RW3 plastic phantom. The beam profiles in *x* and *y* direction have been combined and integrated at all depths to reconstruct a 2D relative dose map. Results for the thick ring are shown in Figure 13.

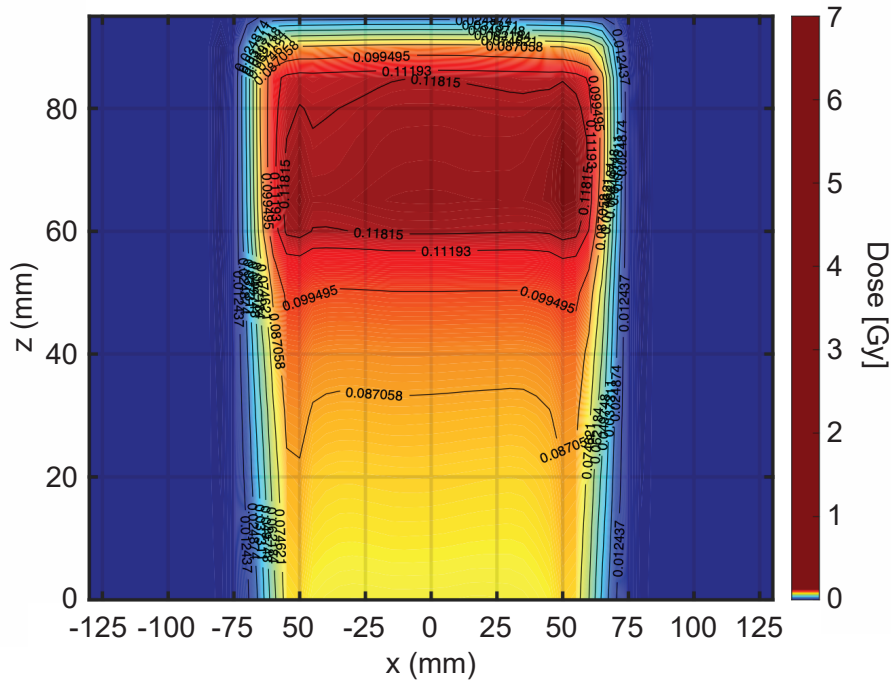

*Figure 13: 2D integrated dose distribution as function of depth in water equivalent (beam direction expressed in z mm), for the thick ring setup. The x (mm) represents the field size around the center of the detector which correspond to the center of the filed.*

## 4. Discussion

This paper shows the extensive work which was performed to establish a versatile R&D proton beam line for research applications. The results show the characterization of the therapeutic pencil beam in terms of Depth Dose Distributions (DDDs), beam shape and size, and beam intensity. Beam energies between 70 and 240 MeV can be delivered at target with energy steps of 0.5 MeV. The single pencil beam spot has a Gaussian shape in both *x* and *y* direction with σ values ranging from 3.6 mm up to 5.4 mm for the maximum and minimum energy, respectively. Beam envelope measurements have been performed for different energies to study the evolution of beam shape and size along the beam path in air. The characteristics of the pencil beam mimic the ones of the clinical Gantry at HollandPTC. It is therefore comparable to the clinical proton beams of other Varian proton therapy centers and its characteristics can easily be compared with clinical proton beam lines in Europe [13].

The beam current express in nA and the proton intensity express in protons per seconds (p/s) were measured with the Faraday cup, to calculate the transmission efficiency (TE) of the cyclotron due to the energy selection system (ESS). A TE of 0.04% was found for the lowest energy resulting in a proton intensity of $2.44 \cdot 10^6$ p/s/nA. A TE of 6% was found for 240 MeV beam energy resulting in a proton flux of $3.3 \cdot 10^8$ p/s/nA. The cyclotron can reach an energy up to 250 MeV, where the degrader of the ESS is fully retracted. This allows the system to achieve a TE of 43%, which allows the production of FLASH dose rates [14] [15]. The characterization of the so-called FLASH beam of the R&D proton beam line will be part of a separate work.
The dual ring passive scattering system was designed following the concept used in the Trento Proton therapy center [11]. Experimental results show field sizes ranging from 2x2 cm$^2$ up to 20x20 cm$^2$ with a uniformity between 97% and 99%. Geant4 Monte

Carlo simulations have been carried out to reproduce and benchmark the passively scattered field and results are presented in [16].

Moreover, the beam was spread longitudinally with the use of a 2D range modulators able to produce a Spread-Out-Bragg peak (SOBP) of 20 mm and 25 mm with a 98% uniformity for thin and thick ring setup, respectively. A maximum dose rate of 9.9 Gy/min at the middle of SOBP can be reached for field sizes up to 8 cm x 8 cm with thin Ring setup, while maximum dose rate of 1.6 Gy/min can be reached with field dimension up to 10x10 cm$^2$ with Thick Ring setup. The field of 20x20 cm$^2$ can reach a maximum dose rate of approximately 0.5 Gy/min at the entrance.

Achieving a dose rate up to ~10Gy/min with a field size of 8x8 cm$^2$ is a great advantage because it allows radiobiological irradiations to be performed in a very short time. This feature allows HollandPTC to conduct radiobiological experiments between patient treatments, allowing researchers to work during the day.

Even though the passive scattering system offers greater flexibility and versatility than other systems available in the world, it will never be comparable to a fully active scanning system. Therefore, for certain applications, HollandPTC offers the possibility of using the clinical Gantry for research. In this way, HollandPTC has both the advantages of a passive scattering system (easy to use, simple geometry, and well-defined dosimetry) and of an active scanning system (fully mimicking patient treatment, possibility to use a treatment planning system, and allowing complex geometries).

In the R&D beam line, the experimental setup with thin and thick ring for in vitro radiobiological experiment with pristine Bragg peak and SOBP is well established and routinely checked with quality assurance procedures. With pre-calibrated setups and characterization, HollandPTC is able to carry out multiple types of experiments ranging from dosimetry, radiobiology, fundamental physics to radioprotection in space. The unique target station provides flexibility for all types of experiments using pencil beam or large fields in a reproducible way.

Since the commissioning of the R&D proton beam line, multiple experiments have been carried out successfully both on dosimetry [17] [18] and radiobiology [19]. Extensive work was carried out to also setup the proton beam line for radioprotection in space applications [20]. These first experiments confirm the versatility of the HollandPTC R&D proton beam line and the usefulness for different types of experiments.

We expect that the beamline will be used extensively in the future to better understand the underlying biological mechanism of proton irradiation and to improve the technology of proton therapy and of areas beyond. Research groups interested in using the facility can submit their proposal to the Research Management Team of HollandPTC (https://www.hollandptc.nl/research/?theme=researchers).

## 5. Acknowledgements

We thank Dr. Francesco Tommasino for providing suggestions on the realization of the first dual ring setup. We thank G.Pittà, M. Lavagno and V. La Rosa of DE.TEC.TOR company to provide support with the devices. We thank Varian Medical Systems (a Siemens Healthineers Company) and the on-site team for their support. This work is part of the PhD thesis of C. Groenendijk, partially funded by Varian.